\documentclass[prl, reprint, superscriptaddress,
               preprintnumbers, amsmath, amssymb, amsfonts]{revtex4-1}

\usepackage{graphicx, dcolumn, bm, here}
\usepackage{psfrag}
\usepackage{epstopdf}
\usepackage{color}
\usepackage{hyperref}

\newlength{\strikewidth} 
\newlength{\strikelength} 
\begin{document}

\preprint{RBRC-945}

\title{Full QED+QCD Low-Energy Constants through Reweighting}

\newcommand{\RBRC}{
  RIKEN BNL Research Center,
  Brookhaven National Laboratory,
  Upton, New York 11973,
  USA}

\newcommand{\UCONN}{
  Physics Department,
  University of Connecticut,
  Storrs, Connecticut 06269-3046,
  USA}

\newcommand{\NAGOYA}{
  Department of Physics,
  Nagoya University,
  Nagoya 464-8602,
  Japan}

\newcommand{\NISHINA}{
  Nishina Center,
  RIKEN,
  Wako, Saitama 351-0198,
  Japan}

\newcommand{\BNL}{
  Physics Department,
  Brookhaven National Laboratory,
  Upton, New York 11973,
  USA}

\newcommand{\INDIANA}{
  Physics Department,
  Indiana University,
  Bloomington, Indiana 47405,
  USA}

\author{Tomomi Ishikawa}
\affiliation{\RBRC}
\affiliation{\UCONN}

\author{Thomas Blum}
\affiliation{\UCONN}

\author{Masashi Hayakawa}
\affiliation{\NAGOYA}
\affiliation{\NISHINA}

\author{Taku Izubuchi}
\affiliation{\RBRC}
\affiliation{\BNL}

\author{Chulwoo Jung}
\affiliation{\BNL}

\author{Ran Zhou}
\affiliation{\INDIANA}

\date{August 17, 2012}

\begin{abstract}
The effect of sea quark electromagnetic charge on meson masses is investigated,
and first results for full QED+QCD low-energy constants are presented.
The electromagnetic charge for sea quarks is incorporated
in quenched QED+full QCD lattice simulations by a reweighting method.
The reweighting factor, which connects quenched and unquenched QED,
is estimated using a stochastic method on $2+1$ flavor
dynamical domain-wall quark ensembles.
\end{abstract}


\maketitle

So far most lattice QCD simulations have been performed neglecting the
electromagnetic (EM) charges.
In order to calculate physical quantities to high precision,
it is quite important to include and control this contribution.
Toward this goal, several attempts regarding this issue
have been done using quenched QED
~\cite{Duncan:1996xy, Duncan:1996be, Blum:2007cy, Blum:2010ym, Torok:2010zz}.

Chiral perturbation theory (ChPT) provides an effective guide to
extrapolate to the physical quark mass point in a lattice calculation,
and in combination with QED, is a powerful tool for addressing
isospin breaking.
The QED effect can be included in the ChPT framework:
partially quenched ChPT (PQChPT) with QED was first derived
by Bijnens and Danielsson~\cite{Bijnens:2006mk} in the $SU(3)$ flavor basis
up to next-to-leading order (NLO) and was recently extended by some of us to 
the $SU(2)$ flavor+kaon basis~\cite{Blum:2010ym}.
In PQChPT, sea and valence quarks are separately treated.
Here, we specify the two valence quarks in mesons by indices $1$ and $3$
and the three sea quarks ($u$, $d$, $s$) by indices $4$-$6$, and
then introduce quark masses $m_i$ and
quark EM charges $q_i$ in units of the fundamental EM charge $e$.
Distinguishing formally the fundamental charges in the sea quark sectors,
$e_{\rm s}$, and in the valence quark sectors, $e_{\rm v}$,
NLO $SU(3)$ PQChPT tells us
the sea EM charge contribution to the pseudo-scalar (PS) meson mass-squared is
\begin{eqnarray}
&&\hspace{-4mm}
\Delta (M_{\rm PS}^{SU(3)})^2
\label{EQ:SU3ChPT}\\
&=&(M_{\rm PS}^{SU(3)}[e_{\rm s}\not=0, e_{\rm v}\not=0])^2
-(M_{\rm PS}^{SU(3)}[e_{\rm s}=0, e_{\rm v}\not=0])^2\nonumber\\
&=&-4e_{\rm s}^2Y_1{\rm tr}Q_{\rm s(3)}^2\chi_{13}
\nonumber\\
&&\hspace{-4mm}
+e_{\rm s}e_{\rm v}\frac{C}{F_0^4}\frac{1}{8\pi^2}
\sum_{i=4,5,6}
\left(\chi_{1i}\ln\frac{\chi_{1i}}{\mu^2}
-\chi_{3i}\ln\frac{\chi_{3i}}{\mu^2}
\right)q_i(q_1-q_3),\nonumber
\end{eqnarray}
with
$\chi_{ij}=B_0(m_i+m_j)$, $Q_{\rm s(3)}={\rm diag}(q_4, q_5, q_6)$.
$\mu$ is an energy scale below which the effective theory accurately
describes the theory (QED+QCD), and $B_0$, $F_0$, $C$ and $Y_1$ are
low-energy constants (LECs).
Determination of $Y_1$ requires $e_{\rm s}\not=0$,
which is not accessible in quenched QED (qQED).
Note that the LECs generally depend on masses and EM charges of 
heavier dynamical quarks than $u$, $d$ and $s$.
In the following, three-flavors of dynamical quarks are assumed:
$u$, $d$ and $s$.
We also mention that a remarkable feature in the three-flavor theory,
\begin{eqnarray}
{\rm tr}Q_{\rm s(3)}=0,
\label{EQ:trQ_2+1flavor}
\end{eqnarray}
makes many terms vanish and leads to the simple form in
Eq.~(\ref{EQ:SU3ChPT}).
Recently, RBC and UKQCD collaborations pointed out that
the $SU(2)$ ChPT is preferable to the $SU(3)$ ChPT
even in the three-flavor full QCD (fQCD) simulation,
since the $s$ quark mass is not small enough~\cite{arXiv:0804.0473}.
In this case, the sea EM contribution to a pion mass-squared, for example,
can be written as
\begin{eqnarray}
&&\hspace{-4mm}
\Delta (M_{\pi}^{SU(2)})^2\label{EQ:SU2ChPT}\\
&=&-4e_{\rm s}^2\left\{Y_1{\rm tr}Q_{\rm s(2)}^2
+Y_1'({\rm tr}Q_{\rm s(2)})^2
+Y_1''q_6{\rm tr}Q_{\rm s(2)}\right\}\chi_{13}
\nonumber\\
&&\hspace{-4mm}
+e_{\rm s}e_{\rm v}\biggl\{\frac{C}{F_0^4}\frac{1}{8\pi^2}
\sum_{i=4,5}
\left(\chi_{1i}\ln\frac{\chi_{1i}}{\mu^2}
-\chi_{3i}\ln\frac{\chi_{3i}}{\mu^2}
\right)q_i\nonumber\\
&&\hspace{+16mm}
+4(\chi_1-\chi_3)\left(J{\rm tr}Q_{\rm s(2)}+J'q_6\right)
\biggr\}(q_1-q_3)
\nonumber\\
&&\hspace{-4mm}
+4e_{\rm s}e_{\rm v}\left(K{\rm tr}Q_{\rm s(2)}+K'q_6\right)
(q_1+q_3)\chi_{13},\nonumber
\end{eqnarray}
where $\chi_i=2B_0m_i$, $Q_{\rm s(2)}={\rm diag}(q_4, q_5)$.
We remark that additional LECs, $Y_1'$, $Y_1''$, $J$, $J'$, $K$ and $K'$
to Eq.~(\ref{EQ:SU3ChPT}) arise due to lack of the property
(\ref{EQ:trQ_2+1flavor}) in the $SU(2)$ case,
and the LECs in Eq.~(\ref{EQ:SU2ChPT}) generally depend on a mass and
an EM charge of the $s$ quark.

\begin{figure}
\begin{center}
\includegraphics[scale=0.47, viewport = 0 0 520 150, clip]
{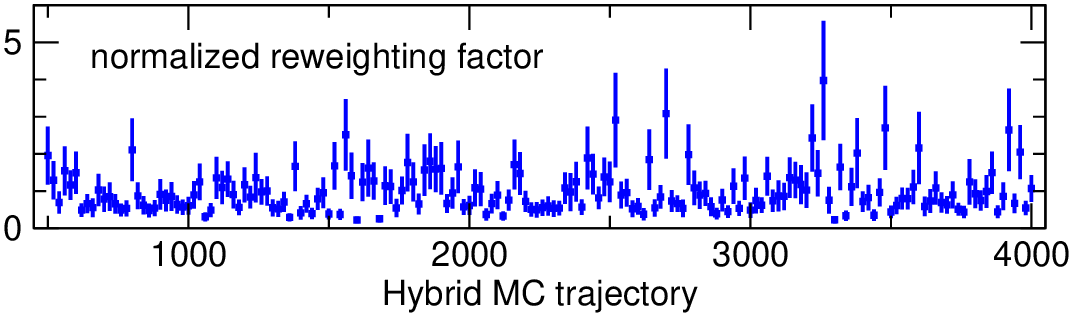}
\vspace*{-5mm}
\caption{Normalized reweighting factor $w[\widetilde{U}, U]$
with the EM charge $e_{\rm s}=e$ on each gluon configuration.}
\label{FIG:normalized_reweight_factor}
\end{center}
\vspace*{-4mm}
\end{figure}

While the full QED (fQED) effect can be incorporated in the Monte Carlo
evolution of the gauge field configuration, the usual gauge ensemble has been
generated only with dynamical QCD.
However, the fQED effect, in principle, can be included
using a reweighting method~\cite{Duncan:2004ys}.
The main purpose of this work is to show the practicality of the reweighting
method for incorporating the sea quark EM charge
on a domain-wall fermion (DWF) ensemble originally generated with $e_{\rm s}=0$
and the feasibility of obtaining the fQED LECs.
(Some applications of reweighting to a realistic QED+QCD simulation
were recently reported in Refs.~\cite{Portelli:2010yn, arXiv:1111.6380}).
Full theory includes
a $U(1)$ photon field $A$ in addition to the usual
$SU(3)$ link variable $U$ for the gluon field and fermion field $\psi$.
In order to illustrate the reweighting method,
we consider the system with a fermion action
$S_f[\bar{\psi}, \psi, \widetilde{U}]=-\bar{\psi}D[\widetilde{U}]\psi$,
where $\widetilde{U}$ is the combined $SU(3)\times U(1)$ gauge link
variable associated with a quark with EM charge $qe$;
\begin{eqnarray}
\widetilde{U}=U e^{iqeA}.
\label{EQ:superimpose}
\end{eqnarray}
Here, we assume the photon fields are generated by
a non-compact $U(1)$ photon action;
\begin{eqnarray}
S_{U(1)}[A]=\frac{1}{4}\sum_x\sum_{\mu,\nu}
\left(\partial_{\mu}A_{\nu}(x)-\partial_{\nu}A_{\mu}(x)\right)^2.
\end{eqnarray}
In this study the fine structure constant of QED is set to be
$\alpha_{\rm EM}=e^2/(4\pi)=1/137$.
An expectation value for some observable $O$ in
fQED+fQCD is formally related to
the one in qQED+fQCD, in which the photon
fields in the quark determinants are neglected, via
\begin{eqnarray}
\langle O\rangle_{\rm fQED+fQCD}
=\frac{\langle wO\rangle_{\rm qQED+fQCD}}{\langle w\rangle_{\rm qQED+fQCD}},
\label{EQ:full-quench}
\end{eqnarray}
introducing a reweighting factor~\cite{Duncan:2004ys},
\begin{eqnarray}
w[\widetilde{U}, U]=\frac{\det(D[\widetilde{U}])}{\det(D[U])}.
\label{EQ:reweight_factor_basic}
\end{eqnarray}
The determinants in Eq.~(\ref{EQ:reweight_factor_basic}) are calculated
by a stochastic estimate with random Gaussian noise vectors.
Since the distribution of $w$ has a long tail, a naive application
of the stochastic estimator for $w$ could fail
~\cite{Ishikawa:2010tq}.
To evaluate $w$ safely, breaking up the determinant into many small
pieces is efficient, because the effects of the outliers are largely suppressed
~\cite{Ishikawa:2010tq, Hasenfratz:2008fg}.
For the splitting, we use a mathematical identity
for the determinant, so called the $n^{\rm th}$-root trick:
$w=\det\Omega=\left(\det\Omega^{1/n}\right)^n$,
which is easily implemented by the rational approximation~\cite{Clark:2006fx}.
We apply reweighting to $2+1$ flavor
dynamical DWF and Iwasaki gluon configurations
generated by the RBC-UKQCD collaborations~\cite{Allton:2007hx}.
The configuration set is one of the ensembles used in the qQED
study~\cite{Blum:2010ym}, whose simulation parameters are 
$\beta_{\rm QCD}=2.13$, $L^3\times T\times L_s=16^3\times32\times16$,
inverse lattice spacing $a^{-1}=1.784(44)$~GeV,
$(am_{\rm u}, am_{\rm d}, am_{\rm s})=(0.01, 0.01, 0.04)$.
The $U(1)$ photon fields, which have been already generated
in the qQED study, are combined with the
gluon configurations according to Eq.~(\ref{EQ:superimpose}).
We also employ $n=24$ roots and use 4 complex random Gaussian noise vectors
per root on each configuration to estimate the reweighting factors.
Fig.~\ref{FIG:normalized_reweight_factor} shows the obtained
factors normalized by the configuration average.
The fluctuation among configurations is moderate, controlled
within a factor of $\sim 5$.

DWF's explicitly break chiral symmetry due to finite size $L_{s}$
in the extra 5th dimension
which can be quantified by an additive, residual, quark mass for each flavor.
In the chiral limit,
$am_{\rm res(QCD)}=0.003148(46)$ for the ensemble used in this study.
The qQED studies~\cite{Blum:2007cy, Blum:2010ym} show that
the valence EM charges further shift the quark mass by an amount of
${\cal O}(\alpha_{\rm EM}am_{\rm res(QCD)})$.
The same effect also arises from the sea quark charges.
This lattice artifact induces a term like
$e_{\rm s}^2\delta_{\rm res}{\rm tr}Q_{\rm s(3)}^2$
in the $SU(3)$ ChPT formula (\ref{EQ:SU3ChPT}).
(Similar modifications are also needed
in the $SU(2)$ formula (\ref{EQ:SU2ChPT}).)
Here we measure the sea EM charge contribution to the residual mass
and subtract it from $\Delta M_{\rm PS}^2$.

Due to finiteness of gauge configurations,
contributions arise from ``hair'', or photon emission to,
and absorption from, the vacuum which averages to zero
in the large ensemble limit.
In Ref.~\cite{Blum:2007cy}, it was shown that this hair is
a large source of noise in hadron correlators.
The leading unwanted piece can, however, be removed by averaging over
plus and minus EM charges, the so-called $\pm e$ trick, 
and it provides a great advantage in which the unphysical noise
is exactly canceled in the valence sector~\cite{Blum:2007cy,Blum:2010ym};
\begin{eqnarray}
\frac{1}{2}\left\{O(+e_{\rm v})+O(-e_{\rm v})\right\}
={\cal O}(e_{\rm v}^2),
\label{EQ:epm-trick_valence}
\end{eqnarray}
where $O(e_{\rm v})$ represents some observable
with a valence EM charge $e_{\rm v}$.
There is also ``hair'' in the sea sector.
To remove the leading contribution from both the sea and valence sectors,
we use an averaging,
\begin{eqnarray}
\frac{1}{2}\left\{O(+e_{\rm s}, +e_{\rm v})
+O(-e_{\rm s}, -e_{\rm v})\right\}
={\cal O}(e_{\rm s}^2, e_{\rm s}e_{\rm v}, e_{\rm v}^2),~~
\label{EQ:epm-trick}
\end{eqnarray}
in the reweighting.
Note that the noise from hair associated with $e_{\rm s}$
is already small by virtue of Eq.~(\ref{EQ:trQ_2+1flavor}).

\begin{figure}
\begin{center}
 \includegraphics[scale=0.42, viewport = 0 0 550 157, clip]
{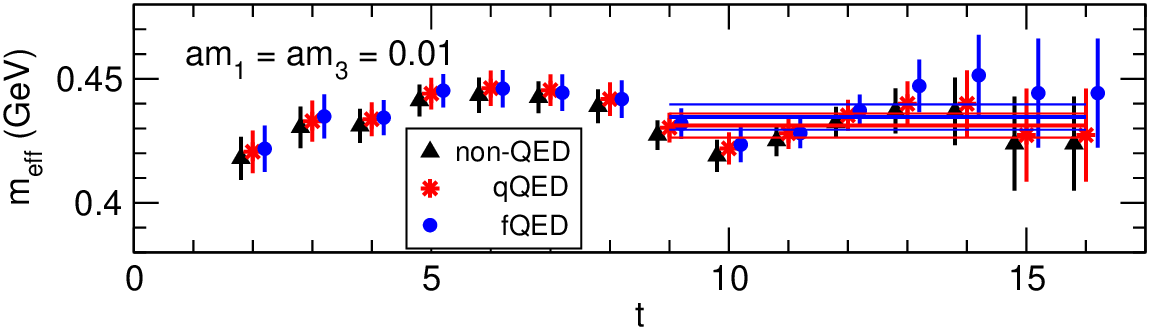}
\vspace*{-3mm}
\caption{
An example of effective mass for the $\pi^+$ meson
in non-QED(black), qQED(red) and fQED(blue)
with $am_1=am_3=0.01$.
The $\chi^2$ fit results of the masses with uncorrelated fit in $t$
are denoted by the horizontal lines.
In fitting the fQED data,
$\chi^2/{\rm d.o.f. ({\bf uncorr})}=0.11$ and
$\chi^2/{\rm d.o.f. ({\bf corr})}=0.67$.}
\label{FIG:effective_mass}
\vspace*{-2mm}
\end{center}
\end{figure}

\begin{figure}
\begin{center}
\includegraphics[scale=0.42, viewport = 0 0 557 161, clip]
{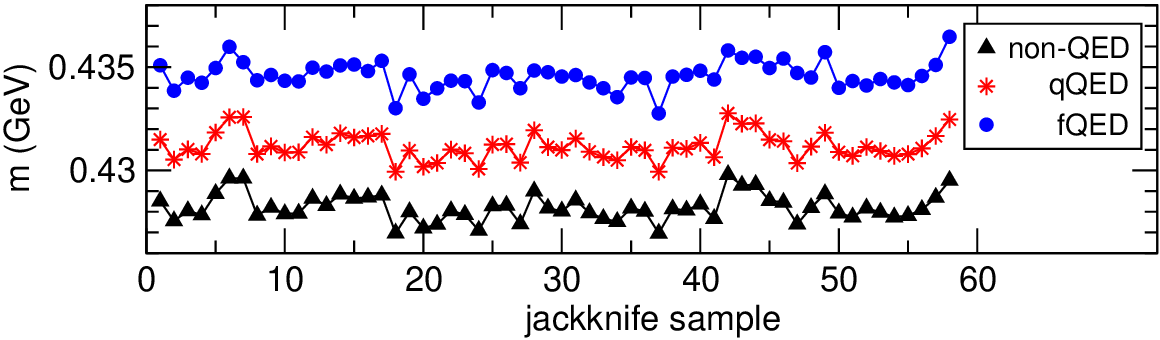}
\vspace*{-3mm}
\caption{
Jackknife data of fit masses of Fig.~\ref{FIG:effective_mass}
({\bf uncorr}).}
\label{FIG:correlation}
\vspace*{-4mm}
\end{center}
\end{figure}

Using the reweighting factor obtained in this work and
the meson correlators in the qQED study~\cite{Blum:2010ym},
the reweighted meson correlators are obtained by Eq.~(\ref{EQ:full-quench}).
An example of effective mass for the $\pi^+$ meson is shown
in Fig.~\ref{FIG:effective_mass}.
For the $\chi^2$ fit results of the masses, we take the same
fit range ($t=9-16$) as in Ref.~\cite{Blum:2010ym}
and also perform both correlated ({\bf corr}) and uncorrelated ({\bf uncorr})
fits in $t$.
(Changing the fit range does not alter results beyond the current statistical
error.)
To study the properties of the data,
we show jackknife samples of fit masses from Fig.~\ref{FIG:effective_mass}
in Fig.~\ref{FIG:correlation}.
Fig.~\ref{FIG:correlation} indicates that the statistical fluctuation comes
mostly from QCD and that significant correlations exist
between the charged and non-charged data.
These facts enable us to detect the qQED and fQED effects.
With the reweighted data of the meson masses calculated,
chiral fits are performed to obtain the QED LECs
in Eqs.~(\ref{EQ:SU3ChPT}) and (\ref{EQ:SU2ChPT}).
Although $C$ is known from the qQED study~\cite{Blum:2010ym},
it provides a valuable consistency check with the qQED result.
In fitting for the LECs, we anticipated a problematic hierarchy
between the $e_{\rm s}^2$ and $e_{\rm s}e_{\rm v}$ terms,
attributable to a double suppression factor in the latter,
\begin{eqnarray}
\frac{m_1-m_3}{m_1+m_3}
{\rm tr}(Q_{\rm s(3)}M_{\rm s(3)})\frac{\overline{m}}{\Lambda_{\rm QCD}},
\label{EQ:supression}
\end{eqnarray}
leaving the  $e_{\rm s}e_{\rm v}$ terms unresolved, where
\begin{eqnarray}
M_{\rm s(3)}=\frac{1}{\overline{m}}{\rm diag}(m_4, m_5, m_6),\;\;
\overline{m}=\frac{m_4+m_5+m_6}{3}.\;\;\;\;\;\;
\end{eqnarray}
Although the difficulty can, in principle, be overcome with enormous
statistics, drastic improvements are provided by engineering
sign flips in the EM charge. Besides the $\pm e$ trick
(Eqs.~(\ref{EQ:epm-trick_valence}) and
(\ref{EQ:epm-trick})), consider a basic transformation
\begin{eqnarray}
{\cal T}_1: (m_1, q_1; m_3, q_3)\longrightarrow (m_3, q_3; m_1, q_1),
\label{EQ:transformaton-1}
\end{eqnarray}
under which the meson system is invariant ($CPT$).
In addition to ${\cal T}_1$, let us introduce transformations:
\begin{eqnarray}
{\cal T}_2: (m_1, q_1; m_3, q_3)&\longrightarrow&(m_1, -q_1; m_3, -q_3),
\label{EQ:transformaton-2}\\
{\cal T}_3: (m_1, q_1; m_3, q_3)&\longrightarrow&(m_3, -q_1; m_1, -q_3).
\label{EQ:transformaton-3}
\end{eqnarray}
Eqs.~(\ref{EQ:transformaton-1})-(\ref{EQ:transformaton-3}) form
a set of transformations that exchange two valence quark
masses and EM charges with, or without, flipping the sign of $e_{\rm v}$.
Note that ${\cal T}_2$ and ${\cal T}_3$ yield only partial invariances
of Eqs.~(\ref{EQ:SU3ChPT}) and (\ref{EQ:SU2ChPT}),
in the sense that the invariance holds only for specific terms in each.
\begin{table}
\caption{
\label{TAB:transformation}
Transformation property
under Eqs.~(\ref{EQ:transformaton-1})-(\ref{EQ:transformaton-3})
for individual terms in NLO $SU(3)$ and $SU(2)$ PQChPT.
}
\begin{ruledtabular}
\begin{tabular}{c|ccc}
               & \multicolumn{3}{c}{terms in NLO PQChPT associated with} \\
transformation & $Y_1$, $Y_1'$, $Y_1''$ & $C$, $\cal J$, ${\cal J}'$ &
$\cal K$, ${\cal K}'$ \\
\hline
${\cal T}_1$ (Eq.~(\ref{EQ:transformaton-1})) & even & even & even \\
${\cal T}_2$ (Eq.~(\ref{EQ:transformaton-2})) & even & odd & odd \\
${\cal T}_3$ (Eq.~(\ref{EQ:transformaton-3})) & even & even & odd \\
\end{tabular}
\end{ruledtabular}
\vspace*{-2mm}
\end{table}
In Tab.~\ref{TAB:transformation}, the transformation property of each term
in NLO PQChPT is summarized.
While the $e_{\rm s}^2$ and $e_{\rm s}e_{\rm v}$ terms retain their even and
oddness under ${\cal T}_1$ and ${\cal T}_2$ to all orders in quark mass,
the transformation property under ${\cal T}_3$ is not preserved
at order higher than $O(am)$ in the quark mass expansion.
At NLO in $SU(2)$ PQChPT in formula (\ref{EQ:SU2ChPT}),
the $e_{\rm s}e_{\rm v}$ term is a mixture of even and odd contributions
since the three-flavor feature (\ref{EQ:trQ_2+1flavor}) is explicitly
broken.
By adding and subtracting squared meson masses related
by these transformations, each term can be separately extracted
and individually fit.
Note that we need at least three different sets of sea quark EM charges
to fully determine the fQED LECs using the $SU(2)$ ChPT;
otherwise we only know their linear combinations (see Tab.~\ref{TAB:LEC}).
A useful choice would be: [${\rm tr}Q_{\rm s(2)}=0$, ${}^{\forall}q_6$],
[${\rm tr}Q_{\rm s(2)}\not=0$, $q_6=0$] and
[${\rm tr}Q_{\rm s(2)}\not=0$, $q_6\not=0$].

\begin{figure*}
\begin{center}
\parbox{85mm}{
\includegraphics[scale=0.42, viewport = 0 0 570 162, clip]
{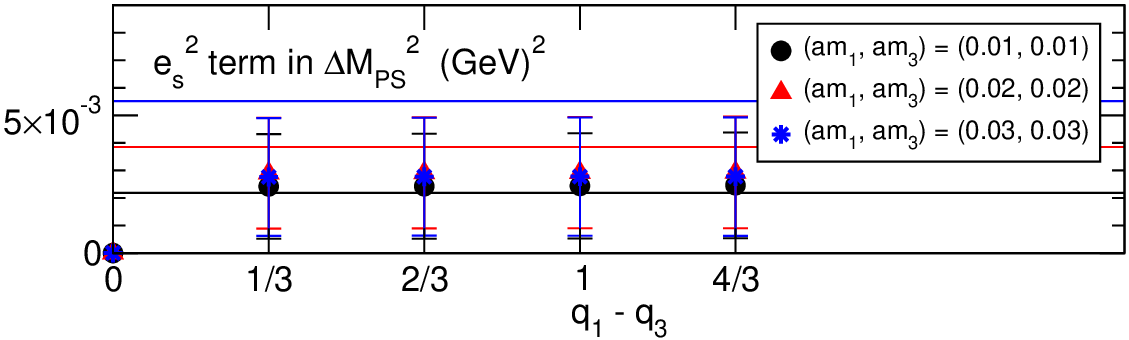}
\vspace*{-6mm}
\caption{
$e_{\rm s}^2$ contribution to $M_{\rm PS}^2$ ({\bf uncorr}).
 Lines represent uncorrelated fits to $SU(2)$ PQChPT.}
\label{FIG:charge_dependence_eses}
}
\hspace*{+2mm}
\parbox{85mm}{
\includegraphics[scale=0.42, viewport = 0 0 570 167, clip]
{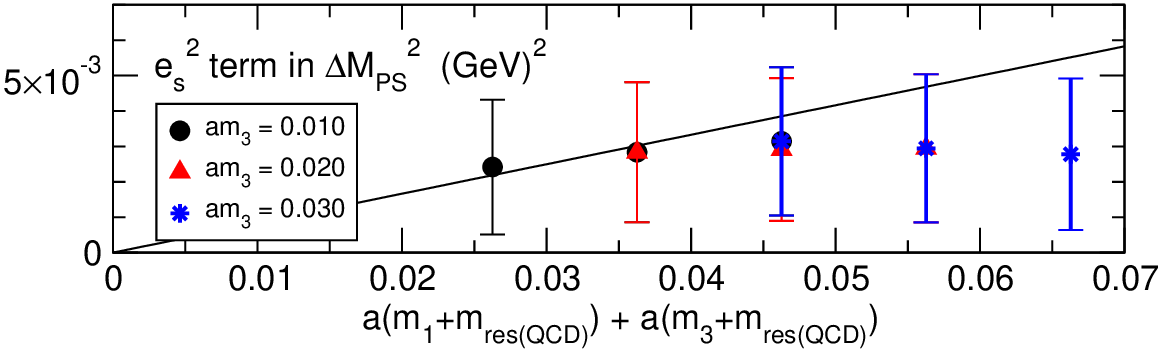}
\vspace*{-7mm}
\caption{
Same as Fig.~\ref{FIG:charge_dependence_eses}
but for $(q_1, q_3)=(+2/3, -1/3)$, showing the valence quark mass dependence.}
\label{FIG:mass_dependence_eses}
}
\end{center}
\begin{center}
\parbox{85mm}{
\includegraphics[scale=0.42, viewport = 0 0 570 162, clip]
{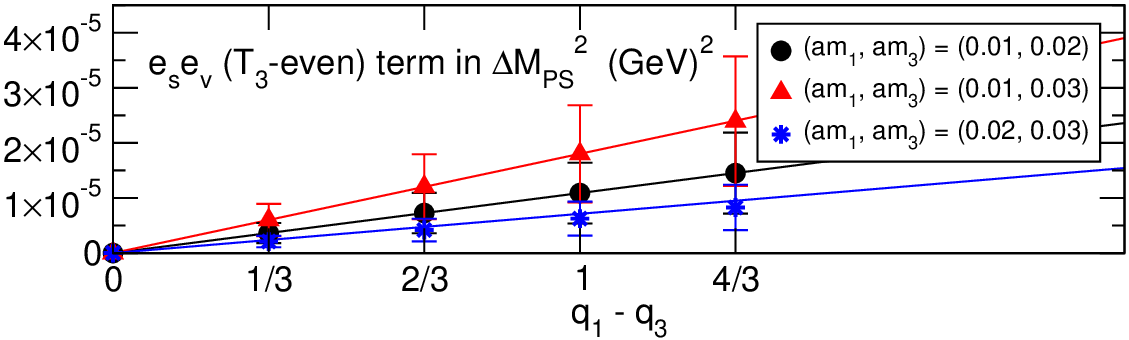}
\vspace*{-6mm}
\caption{
$e_{\rm s}e_{\rm v}$(${\cal T}_3$-even)
contribution to $M_{\rm PS}^2$ ({\bf uncorr}).
Lines represent uncorrelated fits to $SU(2)$ PQChPT.}
\label{FIG:charge_dependence_esev}
}
\hspace*{+2mm}
\parbox{85mm}{
\includegraphics[scale=0.42, viewport = 0 0 570 167, clip]
{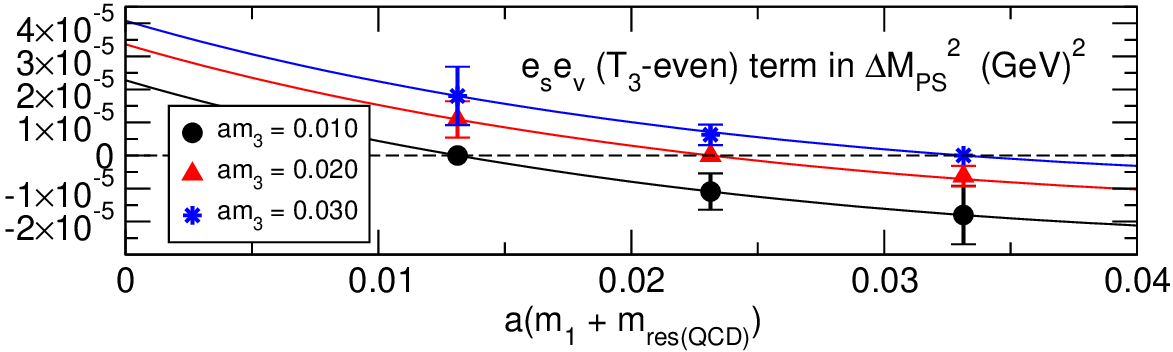}
\vspace*{-7mm}
\caption{
Same as Fig.~\ref{FIG:charge_dependence_esev}
but for $(q_1, q_3)=(+2/3, -1/3)$, showing the valence quark mass dependence.}
\label{FIG:mass_dependence_esev}
}
\end{center}
\begin{center}
\parbox{85mm}{
\includegraphics[scale=0.42, viewport = 0 0 570 162, clip]
{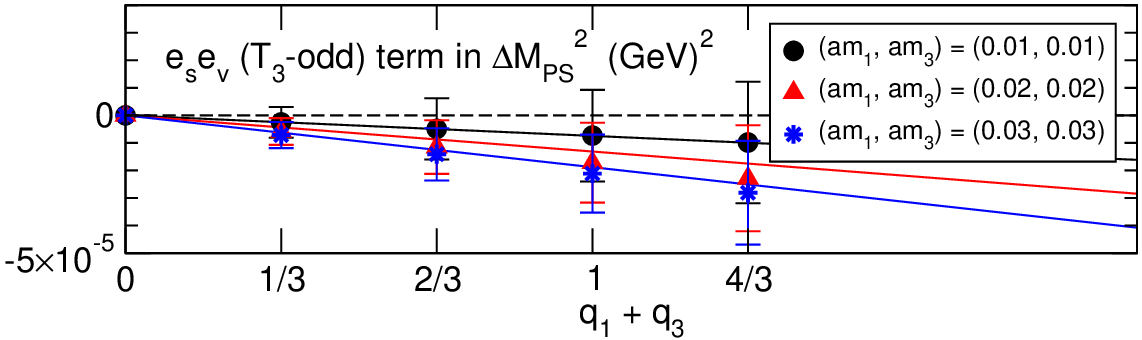}
\vspace*{-6mm}
\caption{
$e_{\rm s}e_{\rm v}$(${\cal T}_3$-odd) contribution to $M_{\rm PS}^2$
({\bf uncorr}).
Lines represent uncorrelated fits to $SU(2)$ PQChPT.}
\label{FIG:charge_dependence_esev_T3odd}
\vspace*{-6mm}
}
\hspace*{+2mm}
\parbox{85mm}{
\includegraphics[scale=0.42, viewport = 0 0 570 167, clip]
{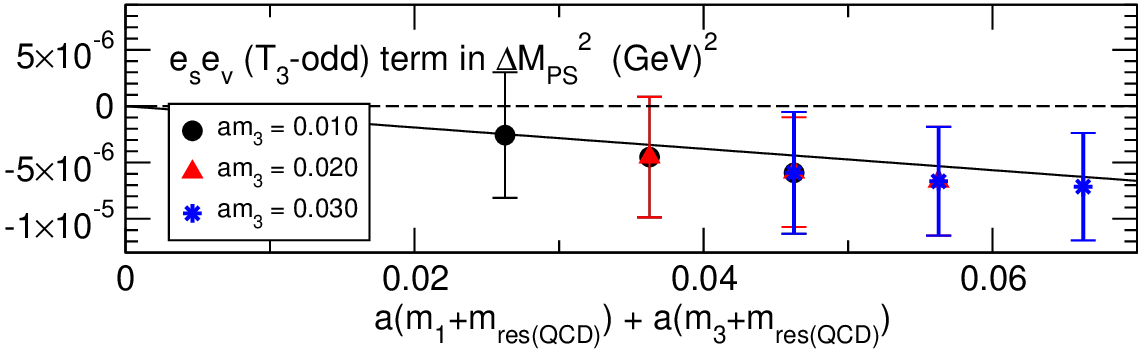}
\vspace*{-7mm}
\caption{
Same as Fig.~\ref{FIG:charge_dependence_esev_T3odd}
but for $(q_1, q_3)=(+2/3, +1/3)$, showing the valence quark mass dependence.}
\label{FIG:mass_dependence_esev_T3odd}
\vspace*{-6mm}
}
\end{center}
\end{figure*}

Figs.~\ref{FIG:charge_dependence_eses}-\ref{FIG:mass_dependence_esev_T3odd}
show individual sea-quark charge contributions to the pion mass-squared,
$e_{\rm s}^2$, $e_{\rm s}e_{\rm v}$(${\cal T}_3$-even) and
$e_{\rm s}e_{\rm v}$(${\cal T}_3$-odd) parts.
The lattice artifact ingredient,
which is caused by the finiteness of $L_s$,
is subtracted from the $e_{\rm s}^2$ term.
In the figures, we can clearly see that
the hierarchy between the $e_{\rm s}^2$ and $e_{\rm s}e_{\rm v}$ terms
is ${\cal O}(10^2)$, as expected
by the suppression given by Eq.~(\ref{EQ:supression}),
and the separation using the transformation ${\cal T}_2$ successfully works.
The valence EM charge dependence is constant for the $e_{\rm s}^2$ term and
linear for the $e_{\rm v} e_{\rm s}$ terms, as expected from the
smallness of the fine structure constant in QED.
\begin{table}
\caption{
\label{TAB:LEC}
QED low-energy constants with $\mu=\Lambda_{\chi}=1$~GeV.
${\cal Y}_1$ is defined as
${\cal Y}_1=Y_1{\rm tr}Q_{\rm s(3)}^2$ for $SU(3)$ ChPT and
${\cal Y}_1=Y_1{\rm tr}Q_{\rm s(2)}^2+Y_1'({\rm tr}Q_{\rm s(2)})^2
+Y_1''q_6{\rm tr}Q_{\rm s(2)}$ for $SU(2)$ ChPT.
$\cal J$ and $\cal K$ depict ${\cal J}=J{\rm tr}Q_{\rm s(2)}+J'q_6$
and ${\cal K}=K{\rm tr}Q_{\rm s(2)}+K'q_6$, respectively.
The qQED values for $C$ are quoted
from Ref.~~\cite{Blum:2010ym}, whose values are obtained
from $24^3\times 64$ lattice and by infinite volume ChPT formula.
The values of $B_0$ and $F_0$ used in the chiral fit are quoted
from Ref.~\cite{arXiv:0804.0473}.
}
\begin{ruledtabular}
\begin{tabular}{c|cccc}
 & \multicolumn{2}{c}{$SU(3)$ ChPT} & \multicolumn{2}{c}{$SU(2)$ ChPT} \\
          & {\bf uncorr} & {\bf corr} & {\bf uncorr} &  {\bf corr} \\ \hline
$10^7C$ (qQED) &  2.2(2.0) &  -- & 18.3(1.8) & --   \\
$10^7C$ &  8.4(4.3) &  8.3(4.7) & 20(14)   &  15(21) \\
$10^2Y_1$ & -5.0(3.6) & -0.4(5.6) & -- & -- \\
$10^2{\cal Y}_1$ & -3.1(2.2) & -0.2(3.4) & -3.0(2.2) & -0.2(3.4) \\
$10^4\cal J$     &  --       & --     & -2.6(1.6) & -3.3(2.8) \\
$10^4\cal K$     &  --       & --        & -3.1(6.9) & -3.7(7.8) \\
\end{tabular}
\end{ruledtabular}
\vspace*{-2mm}
\end{table}
We perform uncorrelated chiral fits for
the $e_{\rm s}^2$, $e_{\rm s}e_{\rm v}$(${\cal T}_3$-even) and
$e_{\rm s}e_{\rm v}$(${\cal T}_3$-odd) terms separately
setting $\mu$ to the chiral scale $\Lambda_{\chi}=1$~GeV
and obtain the LECs in Tab.~\ref{TAB:LEC}.
In this fit, we choose a minimal set of data with smaller valence quark masses,
and ignore $q_6$ dependence in $B_0$ because of smallness of $e^2$ and $Y_1$.
We also neglect finite volume effects
which could give significant shifts in the EM mass spectrum.
However, we remark that our quarks are relatively heavy
even though our lattice is small.
Although the statistical error is large, the value of LEC $C$ is
consistent with that obtained in qQED~\cite{Blum:2010ym}.
(The lattice volume and the quark masses used in the
chiral fit are different between this work and Ref.~\cite{Blum:2010ym}.
The important fact, however, is that the order of magnitude is
consistent between them.)
The size of $Y_1$ seems to be the same
as the other QED LECs in ${\cal O}(e_{\rm v}^2m)$ terms 
determined in qQED~\cite{Blum:2010ym},
which means the sea EM charge effect is comparable to the valence one
except for the Dashen term.

In this study incorporating sea quark EM charges in 2+1
flavor lattice QED+QCD, we have shown that the QED LECs are
accessible using the reweighting method, and that the sea quark LECs are
the same size as the valence ones, as expected.
In our analysis,
the sign flip engineering of EM charges proved to be highly
effective, similar to the $\pm e$ trick
for the valence sector~\cite{Blum:2007cy,Blum:2010ym}.
Since this is a first computation of sea EM charge effects in large scale
computation, our primary aim is to show the method works and the
size of the statistical error.
Checks for systematic errors including the discretization error,
which is a few percent on this lattice for pure QCD~\cite{Aoki:2010dy},
finite volume, and so on, are being investigated on larger lattices,
$24^3\times64$ and $32^3\times64$.
Implementation of further algorithmic improvements,
for example, low-mode averaging to increase statistics, are also in progress.

\begin{acknowledgments}
We are grateful to USQCD and the RBRC for providing computer time
on the DOE and RBRC QCDOC supercomputers at BNL for the computations
reported here.
T.~B and T.~Ishikawa were supported by the U.S. DOE under Grant
No.~DE-FG02-92ER40716.
M.H. is supported by JSPS Grants-in-Aid for Scientific Research
No.~(S)22224003 and No.~(C)20540261.
T.~Izubuchi is partially supported by JSPS Kakenhi grant
No.~22540301 and No.~23105715.
T.~Izubuchi and C.~J are supported by DOE under Contract
No.~DE-AC02-98CH10886.
\end{acknowledgments}


\end{document}